\documentstyle[12pt,a4,epsf]{article}

\begin{document}
\begin{titlepage}
\begin{flushright}
{Revised}\\
{NORDITA-95/36 N,P}\\
{hep-ph/9505251}\\
{Phys. Rev. Lett. 75(1995)1447 + erratum}
\end{flushright}
\vspace{2cm}
\begin{center}
{\large\bf Hadronic Light-by-Light Contribution\\to the Muon $g-2$}\\
\vfill
{\bf Johan Bijnens$^a$, Elisabetta Pallante$^a$
 and Joaqu\'{\i}m Prades$^{a,b}$}\\[0.5cm]
${}^a$ NORDITA, Blegdamsvej 17,\\
DK-2100 Copenhagen \O, Denmark\\[0.5cm]
$^b$ Niels Bohr Institute, Blegdamsvej 17,\\
DK-2100 Copenhagen \O, Denmark
\end{center}
\vfill
\begin{abstract}
We present a calculation of the hadronic light-by-light contributions
to the muon $g-2$ in the $1/N_c$ expansion. We have used an Extended
NJL model and introduced an explicit cut-off for the high energy region.
We have then critically studied the relative size of the high energy
contributions.  Although we find them large we can give a conservative
estimate of the light-by-light contribution to $a_\mu$ which is
 $-(11\pm5) \cdot 10^{-10}$.
This is between two and three sigmas the expected
experimental uncertainty at the forthcoming BNL experiment.
\end{abstract}
\vspace*{1cm}
\vfill
September 1995
\end{titlepage}

The high precision measurement of the
anomalous magnetic moment of the muon, $a_\mu
\equiv (g_\mu - 2)/2$, combined with the LEP
results is expected to provide valuable
information on the
electroweak sector of the Standard Model and maybe
unravel new physics. See \cite{Kinoshita} for reviews
of both theoretical and experimental status.
For carrying out this program, a new measurement of the
muon anomalous  magnetic moment at Brookhaven National
 Laboratory \cite{BNL}
aims to reduce the experimental uncertainty
to $\sim 4 \cdot 10^{-10}$. On the theoretical side the
error is dominated by the hadronic contributions and in
particular by the hadronic vacuum polarization contribution.
For a recent determination of this contribution and
earlier references see \cite{EJ}. The progress expected in measuring
the total cross section $\sigma_{\rm total}(e^+ e^- \to {\rm hadrons})$
will get the theoretical uncertainty from
this contribution down to the order of the experimental
uncertainty quoted above \cite{EJ}.

\begin{figure}
\begin{center}
\leavevmode\epsfxsize=12cm\epsfbox{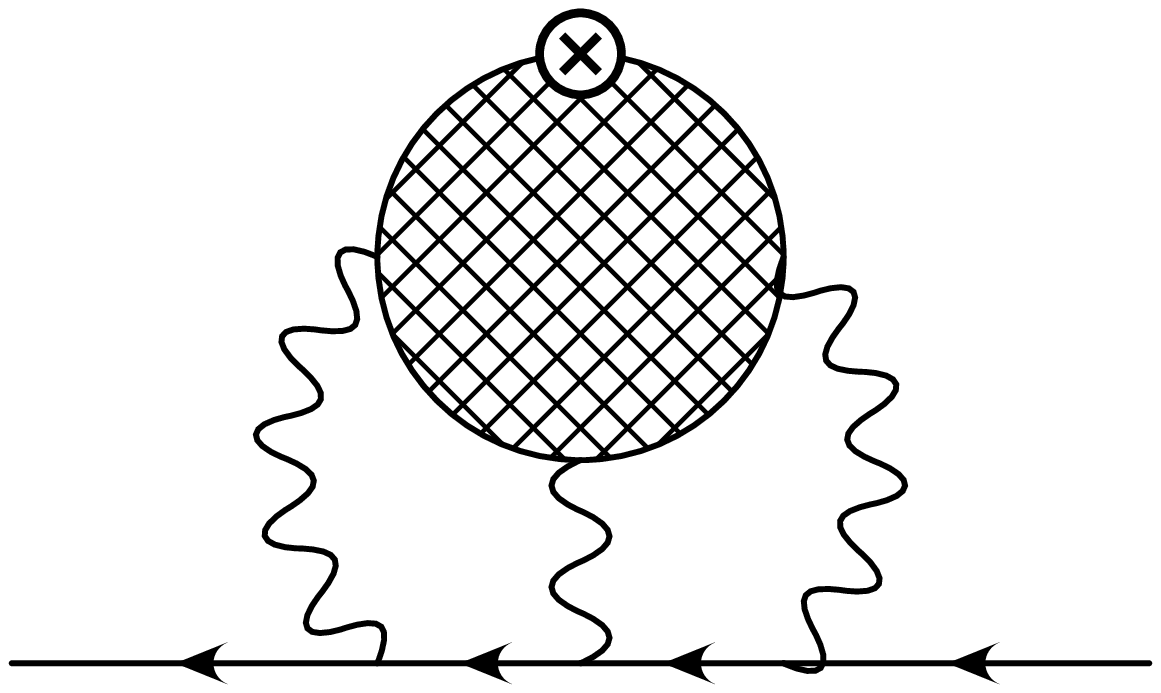}
\end{center}
\caption{}
\label{fig1}
\end{figure}

There is another source of hadronic
uncertainty in the theoretical calculation of $a_\mu$
which has raised
recently some discussion about its reliable
 calculation\cite{BR,Einhorn,deRafael}.
It is the hadronic light-by-light scattering contribution where
a full
four-point function made out of four vector quark currents is
attached
to a muon line with three of its legs coupling to photons
in all possible ways and the fourth vector leg coupled to
an on-shell external photon (see Figure \ref{fig1}).
The difficulty
here is that this contribution cannot be expressed in terms
of experimental observables and thus one has to rely
on our present knowledge in treating the strong interactions.
There have been several attempts to calculate this contribution
in the past\cite{CNPR,KNO} and more
recently in \cite{HKS}.
In this Letter we mainly address the calculation of the
hadronic light-by-light scattering contributions to $a_\mu$
in the large $N_c$ limit ($N_c=$ number of colors).
These are ${\cal O}(N_c)$ in the $1/N_c$ expansion.
An estimate of the effects of the U(1)$_A$ anomaly will be
also included. We will use an
Extended Nambu--Jona-Lasinio (ENJL) model and introduce an explicit cut-off.
We defer a complete discussion
and a more detailed presentation of our results to a forthcoming
publication\cite{BPP}.

The momenta flowing through the
three vector legs of the four-point function attached to the muon
line run from zero up to infinity then covering both the
non-perturbative and perturbative regimes of QCD.
These two different regimes are naturally separated by the
scale of the spontaneous symmetry breaking ($\Lambda_\chi
\simeq 1$ GeV). Above this scale the strong interaction contributions
have to match the perturbative QCD predictions in terms of quarks
and gluons. At this point we would like also to check the current
common wisdom statement that the bulk of the hadronic light-by-light
contributions to $a_\mu$ is determined by the physics around the
muon mass. In fact this was assumed in all previous calculations.
If this was correct we could attempt to make
a pure low energy calculation that would saturate this contribution.
Were the contributions not negligible at some high scale
we would need a more sophisticated model
to calculate the vector four-point  function.
Indeed this appears to be the case from our results.
We shall only be able to give a conservative estimate.

At very low energies (typically below the kaon mass),
 the framework to study the strong interactions is Chiral Perturbation
Theory (CHPT) and the relevant degrees of freedom are the lightest
pseudoscalar mesons ($\pi$, $K$ and $\eta$). However, as emphasized
in \cite{deRafael}, there appear counterterms
in the calculation of $a_\mu$ which are not determined
by symmetry arguments alone.
Instead we will use a low energy model.
We will use an ENJL model  because it possesses
the following features:
it encodes all the chiral constraints and therefore satisfies
all the QCD Ward identities (both anomalous\cite{BP1} and non-anomalous);
it has spontaneous symmetry breaking;
both an $1/N_c$ expansion and a chiral expansion are possible; it
reproduces the low energy phenomenology and the success of
Vector Meson Dominance (VMD) models.
To a large extent, as shown by the Weinberg Sum Rules,
it also has the correct matching with the
high energy QCD behaviour. Models to
introduce vector fields like the Hidden Gauge Symmetry (HGS) model
used
in \cite{HKS} do not always have this good behaviour. These
characteristics were emphasized for $a_\mu$ in \cite{deRafael}.
See \cite{ENJL} for more details,
definitions and technical points about the specific model we are using.

Its major drawback is the lack of confinement. This
can be smeared out by calculating with constituent quarks far off-shell
and color singlet observables. This model has three parameters
plus the current light quark masses . These are fixed as to reproduce the
experimental pion and kaon masses.
The three parameters can be chosen to be the couplings of the spin zero
$G_S$ and spin one $G_V$ four-quark terms in the ENJL model (see
\cite{ENJL} for details) and, since it is a non-renormalizable model,
the cut-off $\Lambda$ of the regularization which we
chose to be proper-time. Although this regulator breaks in general the Ward
identities we impose them by adding the necessary counterterms including those
for the anomaly \cite{BP1}.
The values of the parameters we use are the ones obtained in the first
reference in \cite{ENJL} from a fit to low energy data:
$G_S =$ 1.216, $G_V =$ 1.263 and $\Lambda =$ 1.16 GeV. Then the
constituent quark masses solution of the gap equation
are $M_u = M_d =$ 275 MeV and $M_s =427$ MeV.
The hadronic vacuum polarization contribution was  estimated
within the ENJL model in \cite{deRafael} using a procedure similar
to the one below. The result agreed within about 15\% with
the one in \cite{EJ}.

Let us proceed to the calculation itself.
We calculate to all orders in the
chiral expansion. Notice that this is needed since we
are integrating over three of the vector momenta.
In previous
calculations the lowest order CHPT result was convoluted
with a naive VMD propagator. It is not clear how this
procedure preserves the QCD Ward identities and
the CHPT expansion itself.

\begin{figure}
\begin{center}
\leavevmode\epsfxsize=11cm\epsfbox{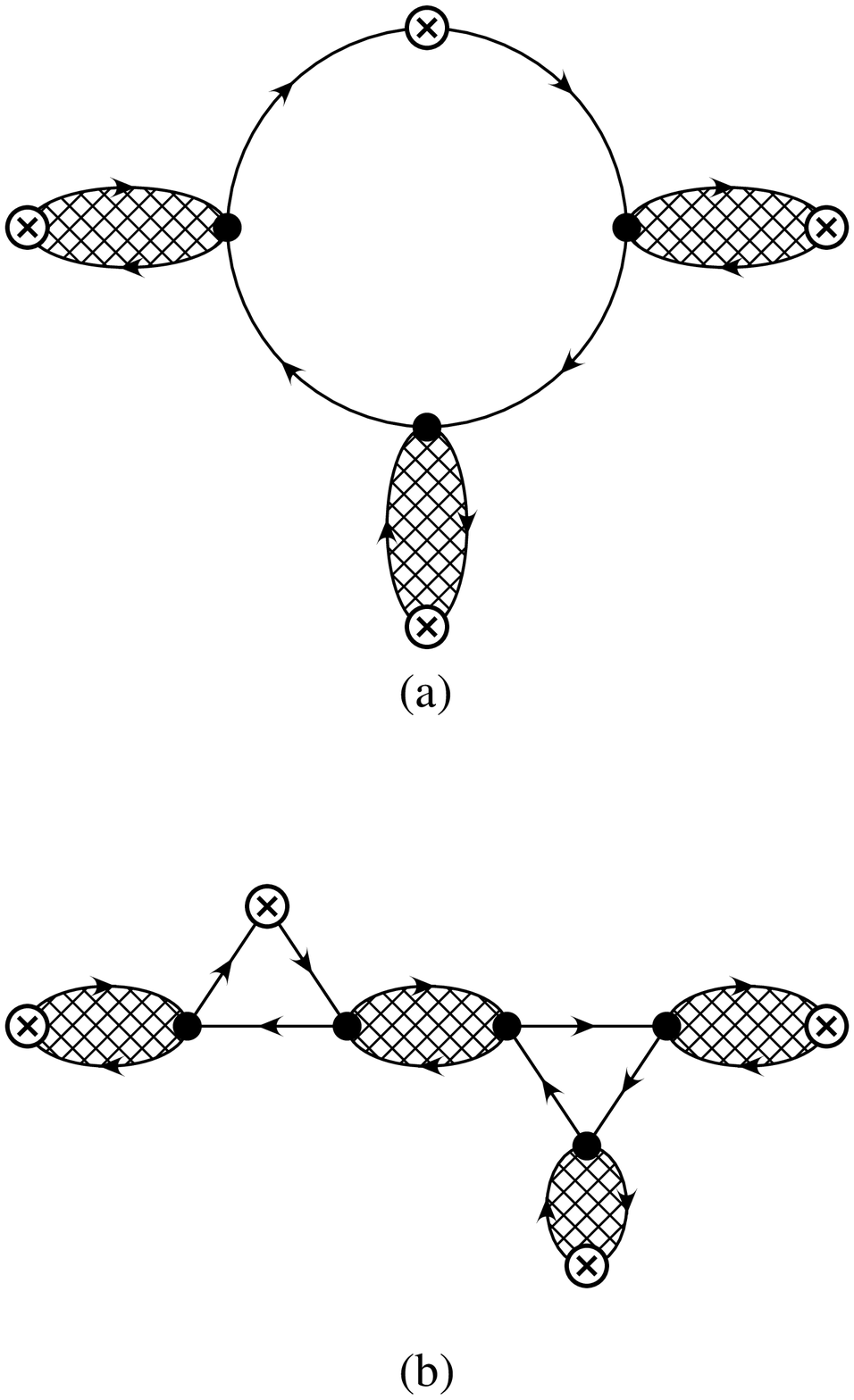}
\end{center}
\caption{}
\label{fig2}
\end{figure}

At leading order in $1/N_c$ there are two
classes of hadronic light-by-light diagrams contributing to $a_\mu$.
The first one is in Figure \ref{fig2}a.
This is a pure full four-point function
with a constituent quark loop and the three vector legs
attaching to the muon line dressed by full two-point functions.
These full two-point functions are the sum of strings with one, two,
$\cdots$, $\infty$
constituent quark loops and can be found in \cite{ENJL}.
There are six
possible permutations for each quark flavor.
The leading order in the CHPT expansion  of this
contribution is ${\cal O}(p^8)$ and thus
potentially sensitive to the high energy region.
The other class is in Figure \ref{fig2}b. Here we have
two one-loop  three-point functions with two vector legs each one
 and glued with a full two-point function
that can be either pseudoscalar, scalar,
mixed pseudoscalar--axial-vector, or  axial-vector.
The vector one does not contribute.
The three vector legs attaching to the muon line are
dressed with full two-point functions.
The leading order in the CHPT expansion is ${\cal O}(p^6)$ for the
pseudoscalar exchange and ${\cal O}(p^8)$ for the others.
There are twelve possible permutations for each quark flavor.

Although the sum of the contributing terms is UV finite, each of
them can be logarithmically divergent and one has to rely
on potentially dangerous numerical cancellations.
 Instead, we used the method proposed in \cite{ABDK} to construct
individually UV safe quantities. This is achieved by making use
of the gauge invariance in the on-shell photon leg. We then construct
the quantity in (2.9) of \cite{ABDK}.
Momentum integrals are performed numerically
in  Euclidean space. This allows us to
impose physically relevant cut-offs on the photons' momenta.

The contribution of the first class of diagrams in Figure \ref{fig2}a
can be written as a seven dimensional integral which we have evaluated
using the Monte Carlo routine VEGAS.
As a check we have reproduced the constituent
quark and muon loops results in \cite{KNO}
and the electron loop results in \cite{LS}.
Since we are dealing with a low-energy model
we want to study the dependence on a high-energy cut-off $\mu$
on the vector legs' momenta.
The result only stabilizes at a rather high value of $\mu$.
For a bare constituent quark with a mass of 300 MeV,
the change between a cut-off of 2 GeV to a cut-off of
4 GeV is still around 20\%.
The change from 0.7 GeV to 2 GeV is typically a factor of 1.8.
This invalidates the use of any low energy
model to calculate the complete
hadronic light-by-light contribution to $a_\mu$.
The bulk of these contributions does not come from the dynamics at scales
around the muon mass as it is often stated. This also explains the rather
high sensitivity to the damping provided by the vector two-point
functions observed in \cite{KNO,HKS}. Although the
only rigorous result is for scales smaller than  (0.6 $\sim$
 1) GeV, one still obtains an estimate.
Mimicking the high energy behaviour of QCD by a bare constituent
quark loop with a   mass of about 1.5 GeV gives only an addition of
$\sim 0.2\cdot10^{-10}$ and if there is any VMD suppression
it will be even smaller. This we will take then
as the uncertainty due to the
high energy region contribution and
the ENJL result where it stabilizes as our estimate.

Let us now turn to the second type of contributions in Figure
\ref{fig2}b. This contribution
can be factorized into a five  dimensional integral which we have evaluated
using the Monte Carlo routine VEGAS times two two-dimensional integrals
and one one-dimensional integral
that we have evaluated using Gaussian integration.
Here, again we have followed the prescription in \cite{ABDK}
to calculate the contribution to $a_\mu$. We have used two different
approaches to calculate the quantity in (2.9) of \cite{ABDK}. The first one
is using the Ward identities for four-point functions and the second
one is using the Ward identities for three-point functions. Both agree exactly.
We have done the same study of the cut-off dependence as for the
four-point function contribution.
The contribution of the pseudoscalar exchange
is more than one order of magnitude larger than the others.
The reason that makes this contribution so different can be
traced back both in the presence of two flavour anomaly vertices and
the CHPT counting.
It therefore deserves more attention. In fact, the pseudoscalar exchange
has  important next-to-leading corrections
from  the effects of the U(1)$_A$ anomaly that
will leave the $\pi^0$ exchange as the dominant contribution to
the muon $g-2$. We have taken into account the effects of the U(1)$_A$
anomaly by using the physical $\pi^0$, $\eta$ and $\eta'$
mass eigenstates  as propagating states.
We see less stability at high values of the cut-off $\mu$ than for
the quark-loop contribution.
Although the change from 0.7 GeV to 2 GeV is also around 1.8, the
stability is worse for cut-off values above 4 GeV.
Notice also that the error from integration routine VEGAS
  is larger for these values
of the cut-off. The poor stability in the pseudoscalar exchange
 is mainly due to the subtraction terms
we need to obtain the correct SU(3) flavour anomaly.
We shall make a detailed analysis of the intermediate and high energy
region contributions to this pseudoscalar exchange in a
more detailed publication \cite{BPP}.

Both scalar and axial-vector contributions are very suppressed
and much smaller than our final error.
In Table \ref{table1} we have listed the ${\cal O}(N_c)$
hadronic light-by-light two leading  contributions
to $a_\mu$ for the up and down quarks as
functions of the cut-off
together with the errors quoted by VEGAS.
We have also included in the fourth column the
estimation of the pseudoscalar exchange  in the presence of the
U(1)$_A$ anomaly. In the fifth column is the sum of the quark
loop and the $\pi^0$, $\eta$ and $\eta'$ exchanges
in the fourth column.
Since the integrand is rather irregular, this error estimate is somewhat
on the small side (see also \cite{Broadhurst}) and will be
largely superseded by the error in our final result.
For the quark loop contribution we used nonet symmetry.
The contribution from the strange quark to the quark loop
is in the range of the quoted errors in Table \ref{table1}.
The charm quark contribution we calculate with a bare quark loop
damped with $c \bar c$ meson dominance propagators in the photon legs.
This contribution is very small.
Both scalar and axial-vector exchange contributions are
again in the range of the quoted errors in Table \ref{table1}.
We therefore take as an estimate of the leading ${\cal O}(N_c)$
hadronic light-by-light contributions to $a_\mu$ including
the effects of the U(1)$_A$ anomaly, the result in the fifth
column of  Table \ref{table1} plus the
scalar and axial-vector exchange contribution,
$0.75 \cdot 10^{-10}$, and the strange and charm quarks contributions,
$0.05 \cdot 10^{-10}$:
\begin{equation}
\label{1}
\left(a_\mu^{\rm light-by-light}\right)_{{\cal O}(N_c)} =
-(10.5\pm5.0)\cdot 10^{-10}\ .
\end{equation}
The error is mainly induced by the uncertainty from the
intermediate and high energy contributions to the
pseudoscalar exchange.

In addition to the leading ${\cal O}(N_c)$ result above there are
the contributions from pion and kaon loops. These
are ${\cal O}(1)$ in the $1/N_c$ expansion and have to be added
to the ${\cal O}(N_c)$ result in (\ref{1}).
We have seen that the lowest order CHPT result is damped
by roughly the same factor in both the constituent
quark loop and the pseudoscalar
meson exchange contributions and that the high energy region
contributes significantly. This can be used to estimate that the
result in \cite{HKS} for the pion and kaon loops is in the right
ball-park when vector mesons are included.
As a first estimate we take the number and error from \cite{HKS}
\begin{equation}
\label{2}
\left(a_\mu^{\rm light-by-light}\right)_{{\cal O}(1)} =
(-0.45\pm0.80)\cdot 10^{-10} \ .
\end{equation}
We will return to this contribution in \cite{BPP}.
Adding the above ${\cal O}(N_c)$ and ${\cal O}(1)$ results
we get our final estimate
\begin{equation}
\label{3}
a_\mu^{\rm light-by-light} = -(11\pm5)\cdot 10^{-10}\ .
\end{equation}

A more general comment is that although a HGS model can be derived
from the ENJL model, this is only true after a series
of approximations. In HGS models the consistency between the
parameters in the anomalous and non-anomalous sectors is not obvious.
In the ENJL model we are using the same parameters appear in both
sectors. This is particularly important
for the flavour anomaly contribution to the light-by-light
scattering. For instance, the calculation
in \cite{HKS} assumes complete VMD for the anomalous sector.
It was shown in \cite{BP1}
that complete VMD breaks the anomalous Ward identities. A prescription
to include vector and axial-vector couplings was given there and was used
in the present work. We find the pseudoscalar-exchange
contribution to be negative and although the central value is
three times larger it is compatible within one sigma
with the values quoted in \cite{HKS}.

Our calculation establishes that
the contribution to $a_\mu$ from light-by-light scattering
is negative and relatively large. It is of the same order
as the one-loop electroweak corrections \cite{oneloop}.
This result is between two
and three sigmas the aimed experimental uncertainty at BNL.
 Our result has a large uncertainty due to intermediate and
high energy contibutions.
Although we believe our estimate is conservative it has
an unsatisfactory uncertainty that will be difficult
to  reduce. We will address this issue in \cite{BPP}.
Despite the uncertainty, the estimate in (\ref{3})
is  still an important theoretical result
for the interpretation
of the muon $g-2$ measurement at the planned BNL experiment.

We thank Eduardo de Rafael for encouragement and discussions.
This work was partially supported by NorFA grant 93.15.078/00.
The work of EP was supported by the EU Contract Nr. ERBCHBGCT
930442.  JP thanks the
Leon Rosenfeld foundation (K{\o}benhavns Universitet) for support and
CICYT (Spain) for partial support under Grant Nr. AEN93-0234.

\newpage

\begin{center}
{\bf TABLE CAPTION}
\end{center}

{\bf Table 1}
\noindent
Results for the two dominant hadronic light-by-light contributions
to $a_\mu$ in the ENJL model.

\begin{center}
{\bf FIGURE CAPTIONS}
\end{center}

{\bf Figure 1}
\noindent
 Hadronic light-by-light contribution to $a_\mu$.
The bottom line is the
muon line. The wavy lines are photons and the cross-hatched circle depicts
the hadronic part. The circled
crossed vertex is an external vector source.

{\bf Figure 2}
\noindent
The two classes of hadronic light-by-light contributions to
$a_\mu$ at leading ${\cal O}(N_c)$. (a) The four-point functions class.
(b) The product of two three-point functions class.
The dots are ENJL vertices.
The circled crossed vertices are where photons
connect. The cross-hatched loops are full two-point functions
and the lines are constituent quark propagators.

\newpage

\begin{table}
\begin{center}
\begin{tabular}{|r|c|c|c|c|}
\hline
Cut-off &  $a_\mu$ $\times$  $10^{10}$ from &
$a_\mu$ $\times$  $10^{10}$ from
&$a_\mu$ $\times$  $10^{10}$ from
& $a_\mu$ $\times$  $10^{10}$ \\ (GeV) &
Constituent &Pseudoscalar& $\pi^0$, $\eta$ and $\eta'$ & \\
&Quark &Exchange ${\cal O}(N_c)$ &Exchanges&  \\
 &  in Figure (2a)  & in Figure (2b) & ${\cal O}(N_c)$ +
U(1)$_A$& Sum \\ \hline
0.7 & $1.14\pm0.02$&$-19.4\pm0.1$ & $-7.2\pm0.1$   & $-6.1$\\
1.0 & $1.44\pm0.03$&$-24.2\pm0.2$ & $-9.4\pm0.1$   & $-8.0$\\
2.0 & $1.78\pm0.04$&$-33.0\pm0.2$ & $-13.2\pm0.2$   & $-11.4$\\
4.0 & $1.98\pm0.05$&$-39.6\pm0.6$& $-15.9\pm0.2$ & $-13.9$\\
8.0 & $2.00\pm0.08$&$-46.3\pm1.5$ & $-18.6\pm0.4$  & $-16.6$\\
\hline
\end{tabular}
\end{center}
\caption{}
\label{table1}
\end{table}

\newpage

\begin{thebibliography}{99}
\bibitem{Kinoshita} Proc. of the 10th Int. Symposium on
``High Energy Spin Physics'', Nagoya, Japan (1992);
Proc. of the Int. Symposium on ``The Future of Muon Physics'',
Heidelberg, Germany (1991), Z. Phys. C56, K. Jungmann,
V.W. Hughes, and G. zu  Putliz (eds.);
``Quantum Electrodynamics'', T. Kinoshita (ed.), World Scientific,
Singapore, (1990) and references therein.

\bibitem{BNL} B.L. Roberts, Z. Phys. C56 (1992) S101.

\bibitem{EJ} S. Eidelman and F. Jegerlehner,
Z. Phys. C67 (1995) 585.

\bibitem{BR} R. Barbieri and E. Remiddi, in ``The DA$\Phi$NE
Physics Handbook'', Vol. II, L. Maiani, G. Pancheri, and N. Paver
(eds.), INFN, Frascati (1992) 301.

\bibitem{Einhorn} M.B. Einhorn, Phys. Rev. D49 (1994) 1668.

\bibitem{deRafael} E. de Rafael, Phys. Lett. B322 (1994) 239.

\bibitem{CNPR} J. Calmet, S. Narison, M. Perrottet, and E. de Rafael,
Phys. Lett. 61B (1976) 283; Rev. Mod. Phys. 49 (1977) 21.

\bibitem{KNO} T. Kinoshita, B. Ni{\u{z}}i\'c, and Y. Okamoto,
Phys. Rev. D31 (1985) 2108.

\bibitem{HKS} M. Hayakawa, T. Kinoshita, and A.I. Sanda,
Phys. Rev. Lett. 75 (1995) 790.

\bibitem{BPP} J. Bijnens, E. Pallante, and J. Prades, in preparation.

\bibitem{BP1} J. Bijnens and J. Prades, Phys. Lett. B320 (1994) 130.

\bibitem{ENJL} J. Bijnens, C. Bruno, and E. de Rafael, Nucl. Phys.
B390 (1993) 501;
J. Bijnens, E. de Rafael, and H. Zheng, Z. Phys. C62 (1994) 437;
J. Bijnens and J. Prades, Z. Phys. C64 (1994) 475;
J. Bijnens, {\em Chiral Lagrangians and Nambu--Jona-Lasinio
like Models},  preprint NORDITA 95/10, hep-ph/9502335 (1995),
to be published in Phys. Rep.

\bibitem{ABDK} J. Aldins, S.J. Brodsky, A.J. Dufner, and T. Kinoshita,
Phys. Rev. D1 (1970) 2378.

\bibitem{LS} B.E. Lautrup and M.A. Samuel, Phys. Lett. 72B (1977) 114.

\bibitem{Broadhurst} P.A. Baikov and D.J. Broadhurst, {\em
Three-Loop QED Vacuum Polarization and the Four-Loop Muon Anomalous
Magnetic Moment}, Open Univ. preprint OUT--4102--54, hep-ph/9504398
(1995).

\bibitem{oneloop}
G. Altarelli, N. Cabibbo, and L. Maiani, Phys. Lett. B40 (1972)
415;
I. Bars and M. Yoshimura, Phys. Rev. D6 (1972) 374;
K. Fujikawa, B.W. Lee, and A.I. Sanda, Phys. Rev. D6 (1972)
2923;
R. Jackiw and S. Weinberg, Phys. Rev. D5 (1972) 2473;
W.A. Bardeen, R. Gastmans, and B.E. Lautrup,
Nucl. Phys. B46 (1972) 319.

\end{thebibliography}
\end{document}